\documentclass[preprint2]{aastex}
\shorttitle{Convection and cluster abundance profiles}
\shortauthors{Rasera et al.}

\def\bm#1{\mbox{\boldmath $#1$}} 
\begin{document}
\title{Abundance profiles in cooling-core clusters: a fossil record of past AGN-driven convection?}

\author{Y. Rasera, B. Lynch, K. Srivastava and B. Chandran}
\affil{Space Science Center, University of New Hampshire, Durham, NH 03824, USA}
\email{yann.rasera@obspm.fr}

\begin{abstract}
Central peaks in the iron abundance of intracluster plasma are a
common feature of cooling-core galaxy clusters. Although centrally localized, 
these abundance peaks have a much
broader profile than the stars of the central brightest cluster galaxy (BCG), which produce
the excess iron, indicating that metal-enriched plasma is transported
out of the BCG by some process such as turbulent diffusion.  At
the same time, cooling-core clusters are likely heated by central
active galactic nuclei (AGNs) by means of jets, cosmic-ray bubbles,
and/or convection. The recent AGN-driven convection model of Chandran
\& Rasera predicts the turbulent velocity profile in a steady-state
cluster in which radiative cooling is balanced by heating from a
combination of AGN-driven convection and thermal conduction.  We use
the velocity profiles from this model as input into an
advection/diffusion model for the transport of metals in the
intracluster medium, taking the iron to be injected by the BCG. 
We compare the results of our model to XMM and Chandra
observations of eight clusters.  Assuming a constant turbulence level
over a cluster's lifetime, the turbulent velocities in the model can
explain the observed abundance profiles in only five of the eight
clusters. However, we go on to develop an analytic fit of the
turbulent velocity profile as a function of the AGN power. We then
deduce for each cluster the average AGN power (during the past $\sim
10$~Gyr) required to match the abundance profiles. The required
average values are between $10^{43}$ and $2\times
10^{44}~\textrm{erg.s}^{-1}$, while the present AGN powers span a much
larger range from $6\times 10^{41}$ (Virgo) to $2 \times
10^{44}~\textrm{erg.s}^{-1}$ (Hydra A). Our results suggest that
AGN-driven convection can account for the observed abundance profiles
if the AGN power varies over a cluster's lifetime between Virgo-like
and Hydra-A-like values, with average values in the above-quoted range.

\end{abstract}

\keywords{galaxies: abundances --- diffusion --- convection --- galaxies: clusters: general --- cooling flows --- galaxies: elliptical and lenticular, cD --- methods: analytical - galaxies: clusters: individual (Perseus, Hydra A, Sersic 159-03, Abell 262, Abell 1795, Virgo, Abell 496, Abell 4059)}

\section{Introduction}

In many clusters of galaxies, the radiative cooling time at the
cluster's center is much shorter than the cluster's age~\citep{fabian94}.
Nevertheless, high-spectral-resolution X-ray observations show that
very little plasma actually cools to low temperatures \citep{bohringer01,david01,molendi01,peterson01,peterson03,tamura01b,blanton03}. This finding, some times referred to as the ``cooling-flow
problem,'' strongly suggests that plasma heating approximately
balances radiative cooling in cluster cores. 

Although a number of different heat sources have been considered in
the literature, there is growing interest in the role of central
active galactic nuclei (AGNs). The importance of AGN heating or ``AGN
feedback'' is suggested by the observation that almost all clusters
with strongly cooling cores possess active central radio sources
\citep{burns90,ball93,eilek04} and by the
correlation between the X-ray luminosity from within a cluster's
cooling radius and the mechanical luminosity of a cluster's central
AGN \citep{birzan04,eilek04}.  One of the main unsolved
problems regarding AGN feedback, however, is to understand how AGN power is
transferred to the diffuse ambient plasma.  A number of mechanisms
have been investigated, including Compton heating
\citep{binney95,ciotti97,ciotti01,ciotti04,sazonov05}, shocks \citep{tabor93,binney95}, magnetohydrodynamic wave-mediated plasma
heating by cosmic rays \citep{bohringer88,rosner89,loewenstein91}, and cosmic-ray
bubbles produced by the central AGN \citep{churazov01,churazov02,reynolds02,bruggen03,reynolds05}, which can heat
intracluster plasma by generating turbulence \citep{loewenstein90,churazov04,cattaneo06} and sound waves
\citep{fabian03,ruszkowski04a,ruszkowski04b} and
by doing $pdV$ work \citep{begelman01,ruszkowski02,hoeft04}.

Another way in which central AGNs may heat the intracluster medium
(ICM) is by accelerating cosmic rays that mix with the intracluster
plasma and cause the ICM to become convectively unstable.  A
steady-state, spherically symmetric model based on this idea was
developed by \citet{chandran04} and subsequently refined by \citet{chandran05}
and \citet{chandran07}.  In this model, a central
supermassive black hole accretes hot intracluster plasma at the Bondi
rate, and converts a small fraction of the accreted rest-mass energy
into cosmic rays that are accelerated by shocks within some
distance~$r_{\rm source}$ of the center of the cluster.  The resulting
cosmic-ray pressure gradient leads to convection (see \citet{chandran06} and \citet{dennis08}), which in turn heats the
thermal plasma in the cluster core by advecting internal energy
inwards and allowing the cosmic rays to do~$pdV$ work on the thermal
plasma. The model also includes thermal conduction and cosmic-ray
diffusion (viscous dissipation turns out to be smaller than the other forms of
convective heating at all radii) and assumes a steady state in which the net heating rate
balances radiative cooling throughout the cluster. The model uses
mixing-length theory to describe convection and its effects on the ICM
and predicts a self-consistent profile for the rms amplitude of the
turbulent velocity.  By adjusting a single parameter in the model
($r_{\rm source}$), \citet{chandran07} were able to achieve a
good match to the observed density and temperature profiles in a
sample of eight clusters.  In several of the clusters in this sample,
compact cooling flows arise within the central few kpc of the clusters
because the rate of radiative cooling peaks much more sharply near the
cluster center than either the convective or conductive heating rates.
At even smaller radii in these clusters, the cooling flow makes a
transition to a Bondi flow, in a manner similar to that described by
\citet{quataert00}. The size of the central cooling flow plays a role
in regulating the mass accretion rate within the central Bondi flow, as
described by \citet{chandran07}.

In this paper, we explore the connection between this
AGN-driven-convection model and the observed properties of the iron
abundance profiles of cooling-core clusters. While observations of
non-cooling-flow clusters show a nearly constant abundance profile,
cooling-core clusters show very peaked iron distributions
\citep{degrandi01}. Observations of the relative abundances of oxygen,
silicon and iron suggest that the production of iron in these
abundance excesses is dominated by SNIa (and possibly winds) from the
central brightest cluster galaxy (BCG) \citep{ettori02,matsushita02,deplaa06}. If cooling
cores are preserved over a timescale longer than $5$~Gyr
\citep{bohringer04}, then the observed amounts of excess iron (of order
$10^8$~M$_{\odot}$) within the cluster core are compatible with the amounts
produced within the BCG by SNIa \citep{cappellaro99} and stellar winds
\citep{ciotti91}.  However, the shape of these abundance profiles is
still a mystery. The distribution of iron is much broader than the
distribution of the stars that produce the metals, which indicates
some additional processes are needed to transport the metals out of
the BCG into the surrounding~ICM \citep{rebusco05}.

Recently, \citet{rebusco05} developed an analytical model of metal
injection by SNIa and diffusion by turbulent gas motions. They
suggested that the dissipation of the same stochastic gas motions
would produce the heating required to solve the cooling-flow problem. Using
this model, \citet{rebusco05,rebusco06,graham06} found that diffusion
coefficients of order $10^{28}-10^{29}~\textrm{cm}^2.\textrm{s}^{-1}$
are required to match the abundance profiles of cooling-core clusters,
while spatial scales $\sim 10$~kpc and velocities of the order of few
$100$~km.s$^{-1}$ are needed to compensate the gas cooling.

Although these studies offered an explanation for the observed
abundance profiles as well as a solution to the cooling-flow problem,
they did not explain how the rms amplitude of the turbulent velocity,
$u$, is determined, or how the dependence of~$u$ on the radial
coordinate~$r$ is determined.  On the other hand, the
AGN-driven-convection model of \citet{chandran07} provides a
physics-based theoretical framework for calculating~$u(r)$.  In this
paper, we combine this model for intracluster turbulence with the
model of metal injection of \citet{rebusco05} in order to
understand the production and transport of metals in the ICM.  We
describe these models further in sections~\ref{sec:model_inj}
and~\ref{sec:model_diff}. In sections~\ref{sec:clusters}
and~\ref{sec:results1} we apply these models to a sample of eight
clusters. We discuss the possible implications of our results for the
variability of the AGN power during a cluster's lifetime in
section~\ref{sec:results2}, and summarize our conclusions in
section~\ref{sec:conclusion}.
 
\section{Model of iron injection by SNIa and winds \label{model}}
\label{sec:model_inj}

From the relative abundance of O, Si and Fe,
\citet{ettori02,matsushita02,deplaa06} have shown that SNIa dominate
the iron enrichment in these abundance peaks. After removing the
contribution from SNII thought to originate from the early formation
of the BCG as well as the background abundance from other
galaxies, the remaining central iron excess originates mainly from
SNIa and stellar winds of the central BCG. Following the work of
\citet{bohringer04,rebusco05, rebusco06}, we estimate the contribution
of SNIa as,
\begin{eqnarray}
\left(\frac{d \rho_{Fe}}{dt}\right)_{SNIa}&=&10^{-12}
\left(\frac{sr}{\rm SNU}\right) \left(\frac{\eta_{Fe}}{\rm M_{\odot}}\right)
 \left(\frac{\rho_{L}}{\rm L^B_{\odot}.kpc^{-3}}\right) {}\nonumber\\ 
 {}&\times& \left(\frac{t}{t_H}\right)^{-k}~{\rm M_{\odot}.yr^{-1}.kpc^{-3}}, \label{eqSNIa}
\end{eqnarray}
with $sr$ the present supernova rate in SNU (supernov{\ae} per century
and per $10^{10}~{\rm L^B_{\odot}}$), $\eta_{Fe}=0.7~{\rm M_{\odot}}$
the iron yield per SNIa, $\rho_L$ the blue luminosity density, t the cosmic time,
$t_H=13.45~{\rm Gyr}$ the current age of the universe and $k$ the exponent
which describes how the supernov{\ae} rate increased in the past. The
wind contribution is taken from \citet{ciotti91,rebusco05},
\begin{eqnarray}
\left(\frac{d \rho_{Fe}}{dt}\right)_{winds}&=&
\gamma_{Fe}\left(\frac{\dot{m}_{wind}}{\rm
  M_{\odot}.yr^{-1}.L^{B-1}_{\odot}}\right) \left(\frac{\rho_{L}}{\rm
  L^B_{\odot}.kpc^{-3}}\right)  {}\nonumber\\ 
 {}&\times&
\left(\frac{t}{t_H}\right)^{-\alpha}~{\rm M_{\odot}.yr^{-1}.kpc^{-3}},\label{eqwinds}
\end{eqnarray}
with $\gamma_{Fe}=2.8 \times 10^{-3}$ the mean iron mass fraction in
the stellar winds of an evolved population, $\dot{m}_{wind}=2.5 \times
10^{-11}~{\rm M_{\odot}.yr^{-1}.L^{B-1}_{\odot}}$ the present star
mass loss rate of a $\simeq 10~{\rm Gyr}$ old population and,
$\alpha=1.3$ specifying the time evolution of the star mass loss
rate. The hidden parameter here is $t_{age}$, which is the age of the
BCG. Eq.\ref{eqSNIa} and Eq.\ref{eqwinds} are indeed only valid
from $t_H-t_{age}$ to $t_H$. Before the formation of the galaxy, the
production of metals is of course assumed to be null.
We assume a \citet{hernquist90} profile typical of elliptical
galaxies,
\begin{eqnarray}
\rho_L(r)&=&\frac{L_B}{2 \pi}\frac{a}{r(a+r)^{3}},
\end{eqnarray}
where $a=r_e/1.8153$ and $r_e$ is the effective radius containing half
of the projected luminosity. 

We constrain the parameters $sr$, $k$ and $t_{age}$ so that the total
observed amount of iron $M_{Fe}$ equals to the total amount of iron
from the model inside the radius corresponding to the last bin of the
abundance observations (located at a radius $r_b$).  Using this
normalization procedure makes the resulting abundance
profile quite insensitive to the particular values of our poorly known
parameters because the only important ingredient for our purpose is
the light profile. For example, varying $k$ between $0$ and $2$ changes
the final abundance by only a few percent (with the diffusion
coefficients suggested by \citet{rebusco05,rebusco06,graham06}).
\citet{renzini93,rebusco06} suggest a value of $k$ between $1$ and
$2$, we pick a value of $k=\alpha$ (that is the time dependence SNIa
iron injection and wind iron injection are the same). In this way, we
factorize the time dependence and adopt a new simple expression,

\begin{eqnarray}
\frac{d \rho_{Fe}}{dt}&=&10^{-12} \times \left(\frac{sr_{eff}}{\rm SNU}\right) \left(\frac{\eta_{Fe}}{\rm M_{\odot}}\right) \left(\frac{\rho_{L}}{\rm L^B_{\odot}.kpc^{-3}}\right) {}\nonumber\\ 
 {}&\times& \left(\frac{t}{t_H}\right)^{-\alpha}~{\rm M_{\odot}.yr^{-1}.kpc^{-3}}\label{eqinj},
\end{eqnarray}
with $\alpha=1.3$ and $sr_{eff}$ an effective supernov{\ae} rate which
includes the wind contribution. This effective supernov{\ae} rate will
be determined directly from the observed mass of iron and is no longer
a free parameter. The advantage of using Eq.\ref{eqinj} is that our
problem of metal production and transport depend now linearly on
$sr_{eff}$. We therefore don't need to run our solver for the
transport equation multiple times to get the right $sr$. Instead, we
run our solver once and then adjust the value of $sr_{\rm eff}$
afterwards (thereby multiplying the abundance profile by a constant) to
match the total observed iron mass.

The last parameter is the age of the BCG, which affects our
results to some degree since an older galaxy allows more time for iron
production and diffusion. A typical galaxy age is given by the stellar
population of the BCG of order $\simeq 8-12$~Gyr
\citep{jimenez07}.  We adopt a fixed age in this range for all our
clusters of $9~$Gyr. The metal enrichment model is now entirely
determined by the observational data $L_B$, $r_e$, $M_{Fe}$.



\section{Diffusion of metals by AGN driven convection}
\label{sec:model_diff} 

We assume that there is steady-state convection in the ICM as
described by the AGN-driven convection model of
\citet{chandran07}. Metals in the ICM are then advected as a passive
scalar by the turbulent flow. The advection equation for the transport
of metals is given by,
\begin{eqnarray}
\frac{\partial \rho a}{\partial{t}}&=&-\bm{\nabla}.(\rho \bm{v} a)+\frac{d \rho_{Fe}}{dt},
\label{eq:psd} 
\end{eqnarray}
with $\rho$ the gas mass density, $a$ the mass fraction of iron, and
$\bm{v}$ the velocity field given by the AGN-driven convection model.
Using mixing length theory and assuming spherical symmetry and a statistical
steady state, we average equation~(\ref{eq:psd}) to obtain the equation
\begin{eqnarray}
&&\frac{\partial <\rho> <a>}{\partial{t}}=\frac{\dot{M}}{4 \pi r^2}
\frac{\partial <a>}{\partial{r}} + <\frac{d \rho_{Fe}}{dt}>{}\nonumber\\ 
 {}&+&\frac{1}{r^2}\frac{\partial}{\partial{r}}\left[r^2 D <\rho>\frac{\partial <a>}{\partial{r}} \right]\label{eqdiff},
\end{eqnarray}
where $\langle \dots \rangle$ indicates an ensemble average, $\dot{M}$
is the mass accretion rate which is independent of the radius and is equal
to the Bondi rate near the center (at $r \approx 0.2$~kpc), $<\frac{d
  \rho_{Fe}}{dt}>\approx \frac{d \rho_{Fe}}{dt}$ is the above metal
production rate from Eq.\ref{eqinj}, $<\rho>=\rho$ is the gas density,
$<a>=a$ is the mass fraction of iron, 
\begin{equation}
D=0.5\,l\, u_{NL}
\label{eq:defD}
\end{equation} 
is the diffusion coefficient, $l=0.4r$ is the mixing length, and
$u_{NL}$ is the radial component of the turbulent velocity from the
mixing-length-theory-based AGN-driven convection model, as calculated
by \citet{chandran07}. The factors 0.5 and 0.4 are choosen to
be consistent with the value used by \citet{chandran07} for the
different energy fluxes. The factor of~$0.5$ in
equation~(\ref{eq:defD}) is much larger than the numerical coefficient
in Equation~8 of \citet{dennis05}, which reads $D_{\rm eddy}
= 0.11 U L \xi$. However, in this last equation, $U$ is the rms value
of the full velocity vector, not just its radial component. Similarly,
$L$ is the full correlation length of the turbulence, whereas the
mixing length~$l$ is just the radial component of the typical
displacement of a fluid element before it is mixed into the
surrounding fluid. (The quantity~$\xi$ accounts for the possible
reduction in~$D_{\rm eddy}$ if the motions are oscillatory and only
weakly turbulent.)  If one sets $u \simeq U/\sqrt{3}$ and $l\simeq
L/\sqrt{3}$, then $D = 0.167 LU$ in equation~(\ref{eq:defD}), which is
similar to equation~8 of \citet{dennis05}. It should also be
noted that if we were to choose a smaller numerical constant in
equation~(\ref{eq:defD}) and throughout the mixing-length model of
\citet{chandran07}, the resulting turbulent velocities in the
model would become larger so that the energy fluxes and convective
heating were still able to compensate for radiative cooling. The
resulting value of~$D$ would not be greatly altered, since it is
closely tied to the rate of convective heating, which is fixed in the
AGN-convection model by the requirement that heating balance cooling
in steady state.

The evolution of the abundance depends on two terms. The first term is
related to the accretion rate $\dot{M}$ and corresponds to an inward
advection of the abundance profile at a speed
$v_{\dot{M}}(r)=\dot{M}/(4 \pi r^2 \rho)$. If we were to assume a
large accretion rate as in the cooling-flow model (e.g.,
$\dot{M}=100-1000$~M$_\odot.$yr$^{-1}$) the accretion of low-abundance
gas from larger radius would destroy the central abundance excess
\citep{bohringer04}.  However, in the AGN-driven convection model,
$\dot{M} \lesssim 1 M_{\sun} \mbox{yr}^{-1}$, and the effect of inflow
is small except within the central few~kpc, where the $1/(\rho
r^2)$~dependence of $v_{\dot{M}}(r)$ causes $v_{\dot{M}}$ to
become~large.  The second term corresponds to diffusion resulting from
the random turbulent velocity field. This term dominates and smoothes
the abundance profile on a scale $r\approx 58 \times
[(D/10^{29}\textrm{cm}^2 \textrm{s}^{-1})\times
  (t_{age}/10\textrm{Gyr})]^{0.5}~\textrm{kpc}$, which is comparable
to the typical half-mass radius of the central iron excesses.

We have solved Eq.\ref{eqdiff} taking a null initial abundance and assuming a
null second derivative of the iron abundance profile for the boundary
conditions. The discretization of the advection part doesn't require any artificial diffusion since the physical diffusion dominates for our case. We have run our code to compute the iron abundance profile of 8 cooling-core clusters.


\section{Properties of the sample of 8 cooling-core clusters \label{sample}}
\label{sec:clusters}

Our sample consists of 8 cooling-core clusters for which we have the turbulent velocity profile from the AGN-driven convection model \citep{chandran07}: Virgo, Abell 262, Sersic 159-03, Abell 4059, Hydra A, Abell 496, Abell 1795 and Perseus. The redshifts of these clusters are taken from \citet{kaastra04}. The properties of these clusters and their brightest cluster galaxies (BCGs) are summarized in table \ref{data} and table \ref{data2}. We use the current $\Lambda$CDM cosmology ($H_0=70$~km.s$^{-1}$.Mpc$^{-1}$, $\Omega_m=0.3$ and $\Omega_\Lambda=0.7$).
 
\subsection{Photometric properties} 

The photometric properties we are interested in are the total luminosity $L_B$, and the effective radius $r_e$, which contains half of the projected flux. We get the name of the brightest cluster galaxies (BCG) in each cluster from \citet{rafferty06} and the NED database, and then compute the total blue luminosity using the Hyperleda database \citep{paturel03} proceeding as follows. We take the apparent corrected blue magnitude of the BCGs, compute the luminosity distance, and the associated distance modulus from our redshift and the $\Lambda$CDM cosmology. We finally compute the blue absolute magnitude, and the blue luminosity assuming $M^B_\odot=5.47$. 

Finding the effective radius is more challenging. We use three different
sources: \citet{schombert87} (Perseus, Abell 1795, and Virgo),
\citet{graham96} (Abell 262 and Abell 496) and Hyperleda (Hydra A, Sersic
159-03, and Abell 4059). In these references an R$^{1/4}$ law
\citep{devaucouleurs53} is adjusted to the observed profile from which the
best fit effective radius (or the effective surface brightness in Hyperleda)
is deduced. We also convert to our cosmology using the ratio of angular
diameter distance for our cosmology and redshift to the angular diameter
distance for the authors' cosmology and redshifts. The results for $r_e$ and
$L_{cr}$ are shown in table \ref{data}. 

We note that we neglect the cores observed on small scales \citep{carollo97}, the deviation from the R$^{1/4}$ law \citep{graham96} on intermediate scale, the large envelopes observed on large scales \citep{schombert88} and the ellipticity of the profiles. We do not hide here that the uncertainties on $r_e$ and $L_{B}$ are important. Fortunately, the result on the shape doesn't depend on $L_{B}$ (due to our renormalization procedure) and has only a little dependency on $r_e$ for most of the clusters because the diffusion scale is larger than $r_e$.

\subsection{Gas density and turbulent velocity profile \label{diffusion}}

\begin{figure}[h!]
  \plotone{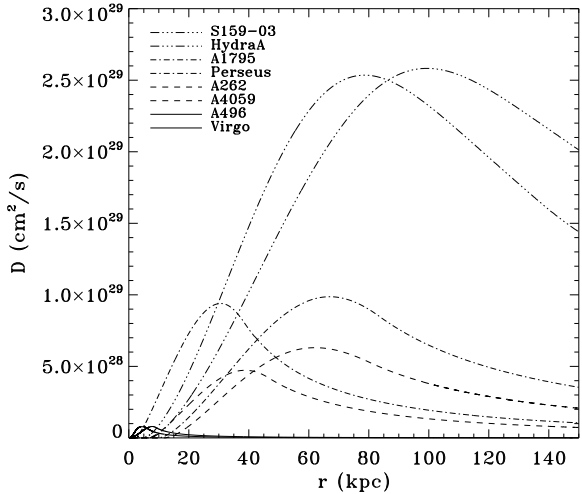}
  \caption{Diffusion coefficient (due to the stochastic turbulent motions of the gas) for our 8 clusters. This comes from the AGN-driven convection model of \citet{chandran07}.\label{dofr}}
\end{figure}

We take the eight clusters in our sample to have the density profiles
and (as discussed above) turbulent-velocity profiles obtained from the
AGN-driven convection model of \citet{chandran07}. The model density
  profiles are very similar to the observations of \citet{kaastra04},
  as shown in Figure~1 of \citet{chandran07}.  We note that
the AGN-driven-convection-model solutions were obtained by varying the
size of the cosmic-ray acceleration region in the model in order to
achieve a best-fit to the observed density and temperature profiles of
the eight clusters in our sample.  The model solutions were therefore
not fine-tuned to solve the independent problem we are investigating
now; the shapes of the abundance profiles. From the turbulent radial
velocity profile $u_{NL}$, we deduce the diffusion coefficient of
Eq.\ref{eqdiff}, $D(r)=0.2 r\times u_{NL}$.

The profiles of the diffusion coefficients are presented in
Fig.\ref{dofr}. The diffusion coefficients peak around a radius
$r_{max}$ where the diffusion reaches its maximum $D_{max}$. We divide
the clusters into 4 levels of diffusion. Virgo and Abell 496 have the
smallest diffusion coefficients, with $D_{max}\approx 8 \times
10^{27}\textrm{cm}^2.\textrm{s}^{-1}$. They are the ``very weak
feedback clusters''. Abell 4059 and Abell 262 have small diffusion
coefficients ($D_{max}\approx 5 \times
10^{28}\textrm{cm}^2.\textrm{s}^{-1}$) and are called ``weak-feedback
clusters''.  Perseus and Abell 1795 have intermediate diffusion
coefficients ($D_{max}\approx 10^{29}\textrm{cm}^2.\textrm{s}^{-1}$)
and are called ``moderate feedback clusters''.  Finally, Hydra A and
Sersic 159-03 have large diffusion coefficients ($D_{max}\approx 2.5
\times 10^{29}\textrm{cm}^2.\textrm{s}^{-1}$) and are the
``strong-feedback clusters''. We summarize the value of $D_{max}$ and
$r_{max}$ in table \ref{data2}.


\subsection{Observed abundance profiles}

\begin{figure}[h!]
  \plotone{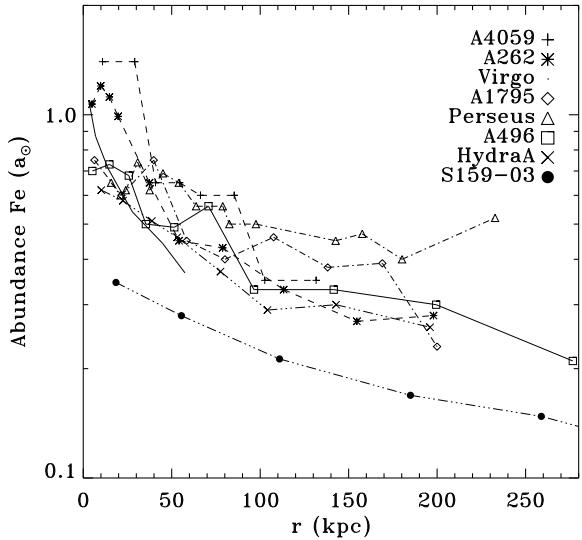}
  \caption{Observed abundance profiles in solar unit for our 8 clusters (Chandra or XMM). We decompose them into a central iron excess from the BCG ($r<100$~kpc) and an asymptotic abundance from the background galaxies ($r>100$~kpc).\label{abundance}}
\end{figure}

We find the observed abundance profiles in the literature (helped in this way
by the BAX cluster database \citep{sadat04}). All the data points come either
from Chandra or XMM observations. We converted all the radii to~kpc using the
angular diameter distance from $\Lambda CDM$ cosmology. Finally, we converted 
all the different solar abundances to the value published by
\citet{anders89} (where the solar abundance relative to H is $4.68\times
10^{-5}$ in number). 
The references used are the following: Virgo \citep{matsushita02},
Abell 262 \citep{vikhlinin05}, Sersic 159-03 \citep{deplaa06}, Abell
4059 \citep{choi04}, Hydra A \citep{david01}, Abell 496
\citep{tamura01a}, Abell 1795 \citep{ettori02}, and Perseus
\citep{churazov03}. We include all the observational points with two
exceptions. We neglect the abundance hole observed within the central
several kpc in Perseus (3 points at $r<10$~kpc), Abell 262 (3 points
are $r<10$~kpc) and Virgo (1 point $r<1$~kpc). In the case of Perseus,
\citet{churazov03} mentions that the profile may be affected by the
bright compact source in this region. We also removed the data points
that suddenly drop to very low values at large~$r$ (the last point in
the case of Sersic 159-03 and two last points in the case of Abell
262). We also note that we used the deprojected profiles for clusters
Perseus, Virgo, Abell 496 and Abell 4059 and projected profiles for
clusters Hydra A, Sersic 159-03, Abell 262 and Abell 1795 because
deprojected profiles were not available. We do not expect this to have
a large effect on our conclusions, because for the cases for which we
have both projected and deprojected profiles, the two types of profile
are similar, with the main difference being that the deprojected
profile is more noisy. The resulting abundance profiles are shown in
Fig.\ref{abundance}.

This figure shows that the ranges of values for the abundance profiles
are relatively similar. We note that the normalization of the
abundance profile of Sersic 159-03 is lower than the others. This
lower normalization may be related to the soft X-ray excess, which is
most likely of non-thermal origin \citep{werner07} and may contribute
to the continuum. For $r<100$~kpc a central abundance peak 
 is observed for all eight clusters while at larger radii
($100-300$~kpc), the profiles flatten and reach an asymptotic value of $a
\sim 0.2-0.4$. This similar shape for all the clusters may indicate a
common mechanism for creating the abundance profiles and allows us to
decompose the profiles in two parts: a flat contribution and a central
excess. We note that three out of the four highest peaks (Virgo, Abell
4059 and Abell 262) correspond to the clusters with less diffusion
(see Fig.\ref{dofr} in our model).


Following the work of \citet{bohringer04,rebusco05, rebusco06}, we
subtract a constant value $a_b$ from the observed abundance
profiles. $a_b$ corresponds to the contribution of iron from the
background galaxies as well from SNII. Our goal is to isolate the
central iron excess or more precisely the iron contribution from the
BCG itself.  Indeed, \citet{degrandi01} measured the abundance
profiles at large radii for 17 clusters. They showed that the
abundance profiles of all non-cooling-core clusters are consistent
with being constant with radius at a value of order 0.2-0.4. On the
other hand, cooling-core clusters present a peak near the center and
an asymptotic value at large radii at the same value of order 0.2-0.4.
Moreover, an analysis of the elemental abundance patern in M87
\citep{finoguenov02} and Sersic 159-03 \citep{deplaa06} has shown that
SNIa dominate the metal enrichment in the central region and that
their contribution decreases with radius. These observations support
the idea of a background value from SNII in galaxies throughout the cluster.

To estimate these background values $a_b$, we take the average values of the
last three points in the vicinity of 100~kpc. Then, to avoid negative abundances, we remove the points were $a<a_b$. Since the observations of Virgo do not reach $100$~kpc and seems to be below the other curves at intermediate radii, we choose a typical value $a_b=0.2$. The last observational points are located at a distance $r_b$ from the center. The values of $a_b$ and $r_b$ are shown in table \ref{data2}.  

From the density profile and the abundance profile, we deduce the total
observed mass of iron in the central excess $M_{Fe}$ (see table
\ref{data}). This mass is the mass of iron inside $r_b$. In our model, we
always choose $sr_{eff}$ so that the mass of iron inside $r_b$ is also equal
to $M_{Fe}$. We also characterize the width of the abundance profile by computing
$r_{1/2}$, the radius which contains half the total observed mass of iron in
the central excess (inside $r_b$).

\section{Impact of present AGN-driven convection on iron distribution}
\label{sec:results1}

\begin{figure*}[h!]                               
   \begin{tabular}{cc}
    \includegraphics[width=0.45\hsize]{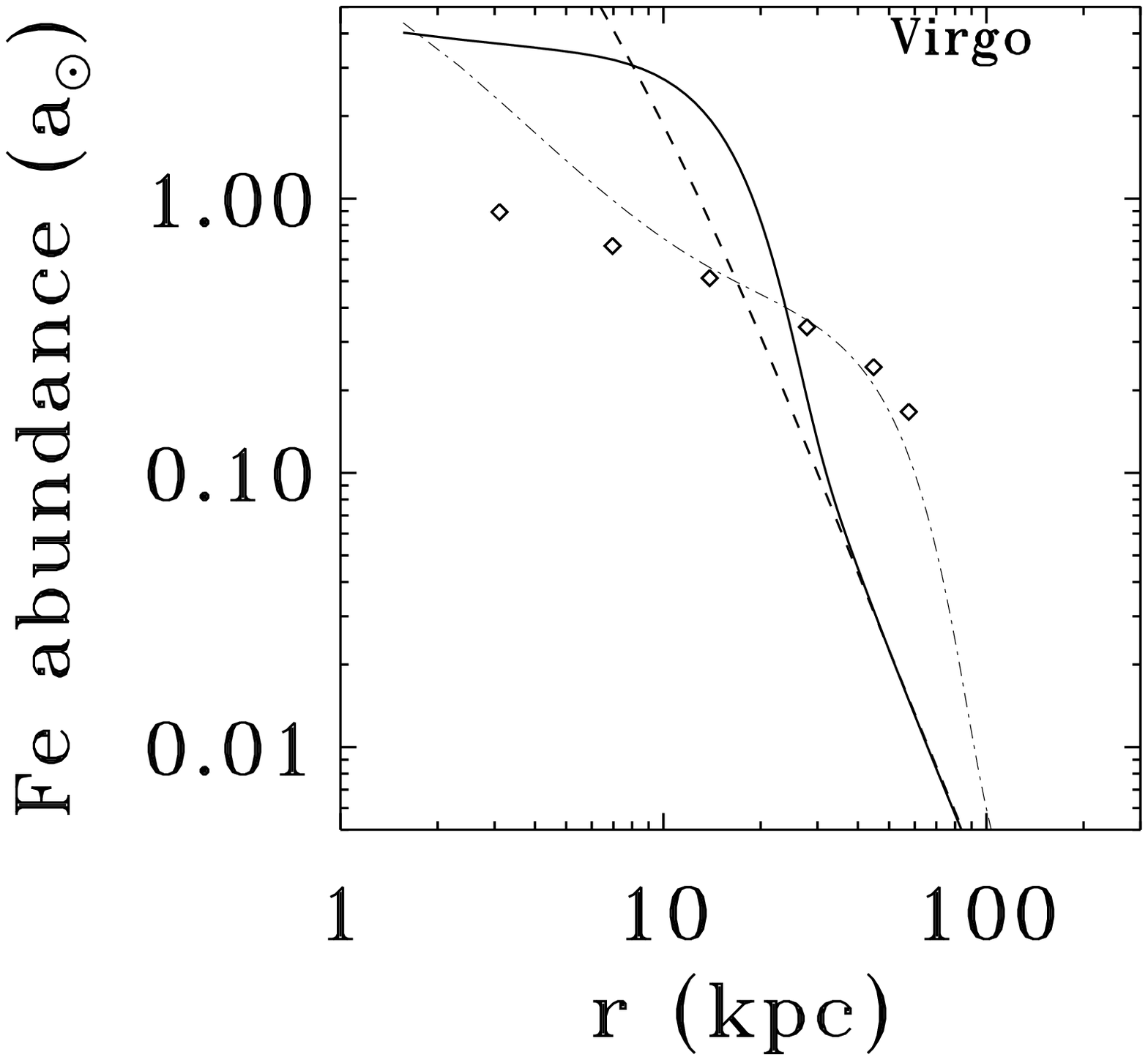}
    \includegraphics[width=0.45\hsize]{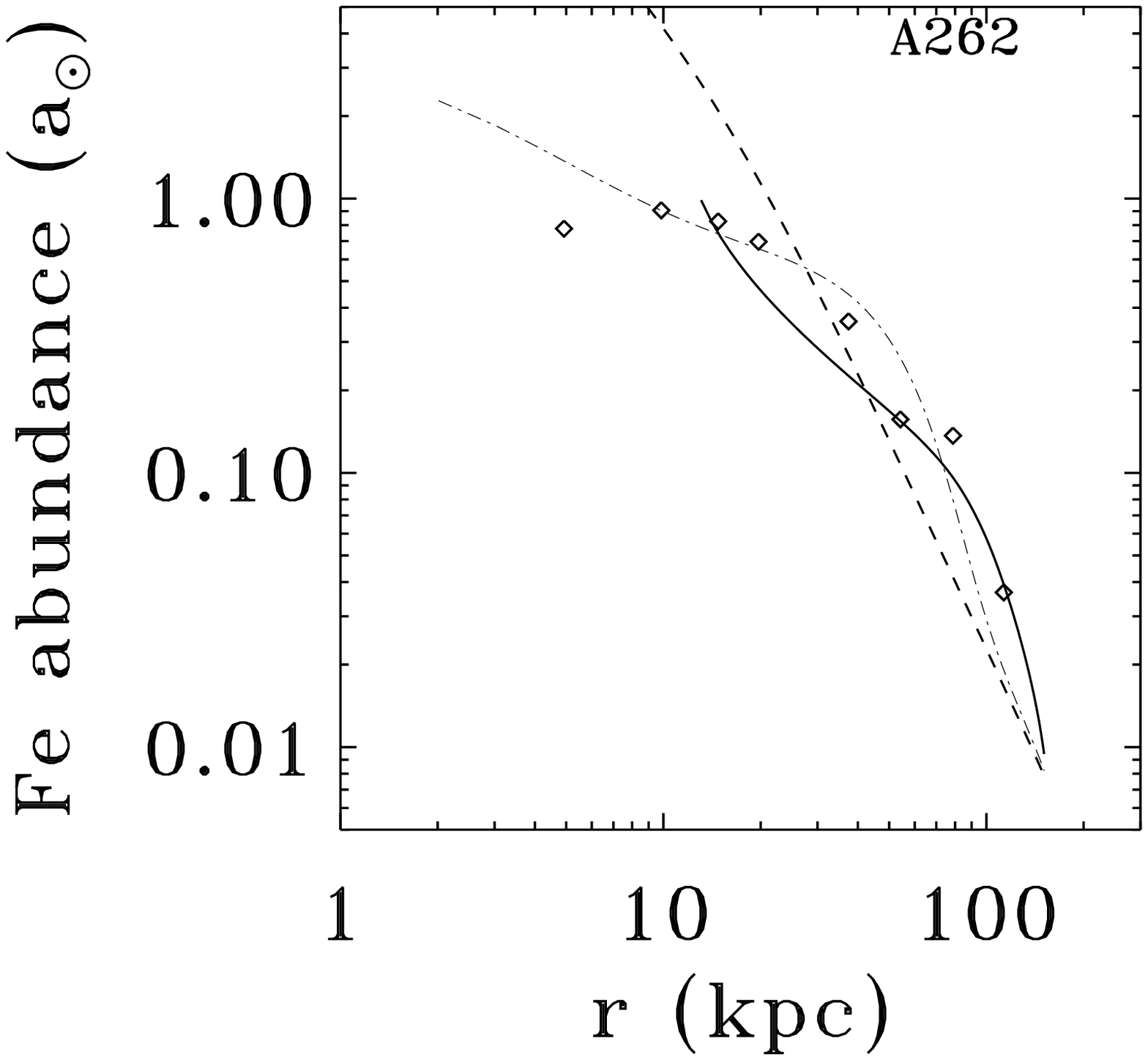}    \\          
    \includegraphics[width=0.45\hsize]{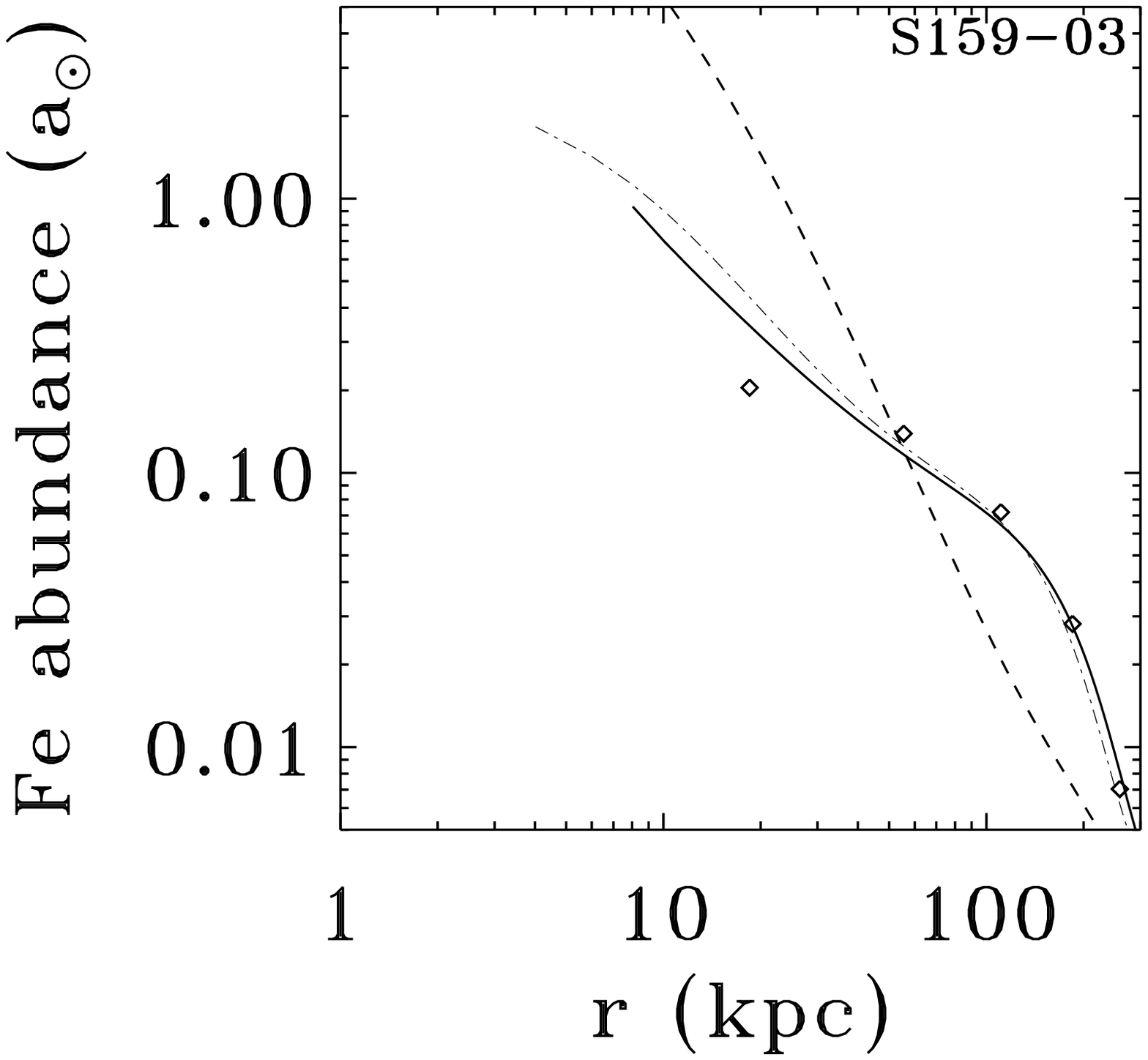}
    \includegraphics[width=0.45\hsize]{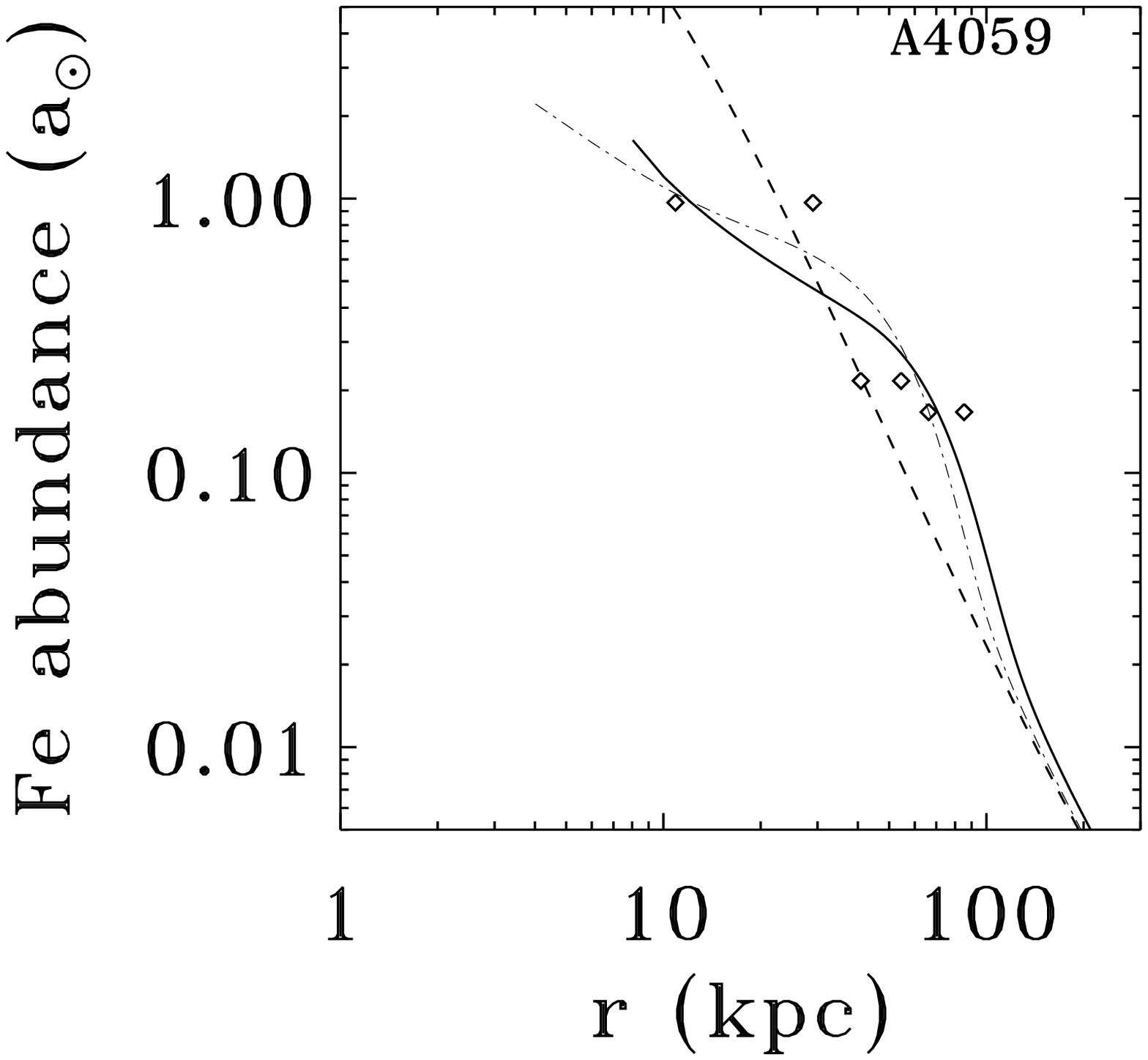}      
  \end{tabular}  
  \caption{Abundance profile in solar unit as a function of the radius in kpc for Virgo, Abell 262, Sersic 159-03, and Abell 4059. The dashed line is the resulting abundance profile without diffusion. The continuous line is the resulting profile with the present cluster diffusion. The dot-dashed line is the profile with the best fit cosmic-ray luminosity value and subsequent diffusion. The diamonds are the observations. Past and present AGN-driven convection seems to play an important role for the transport of iron. \label{result1}} 
\end{figure*}

\begin{figure*}[h!]                           
       \begin{tabular}{cc}
    \includegraphics[width=0.45\hsize]{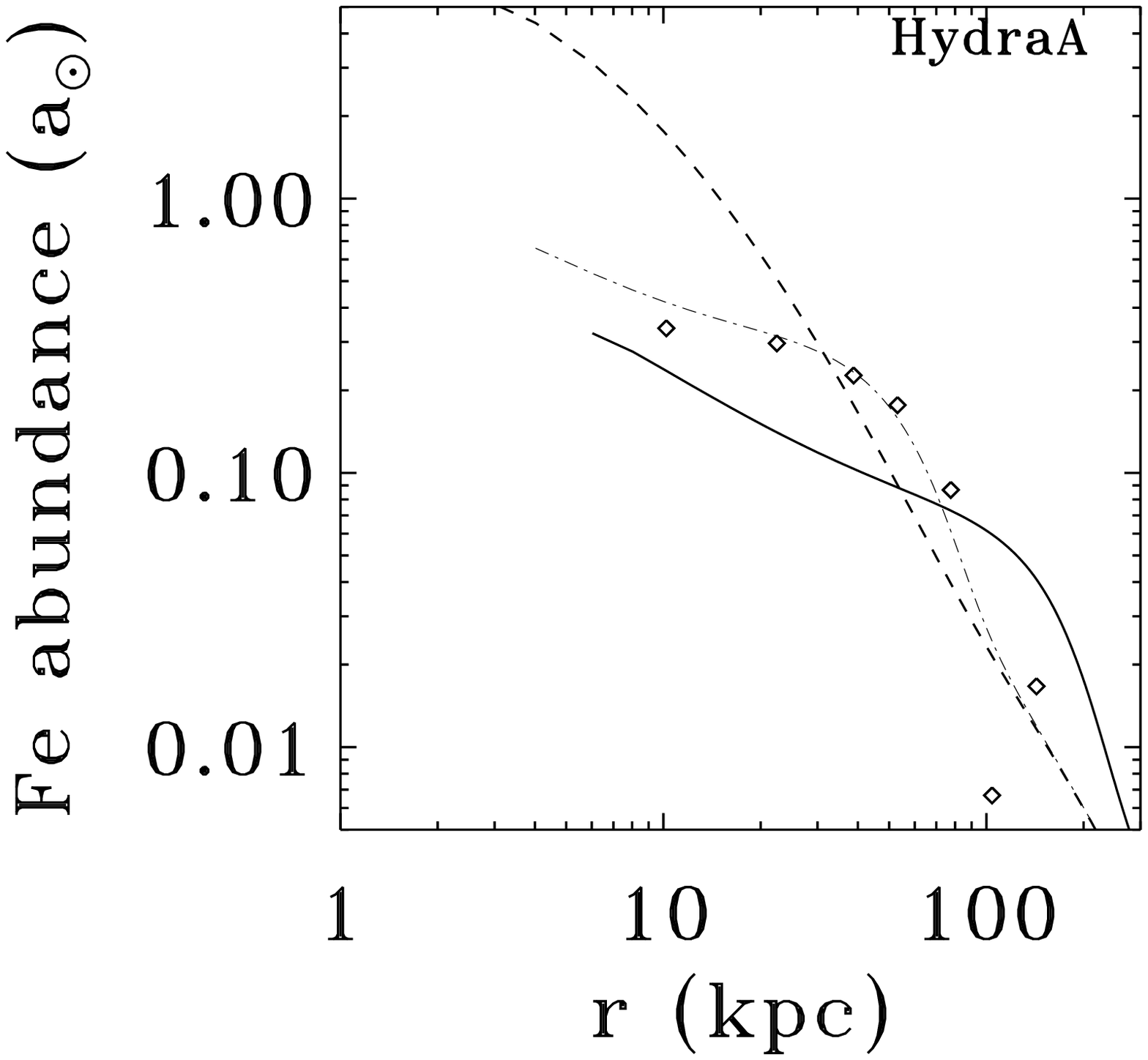}
    \includegraphics[width=0.45\hsize]{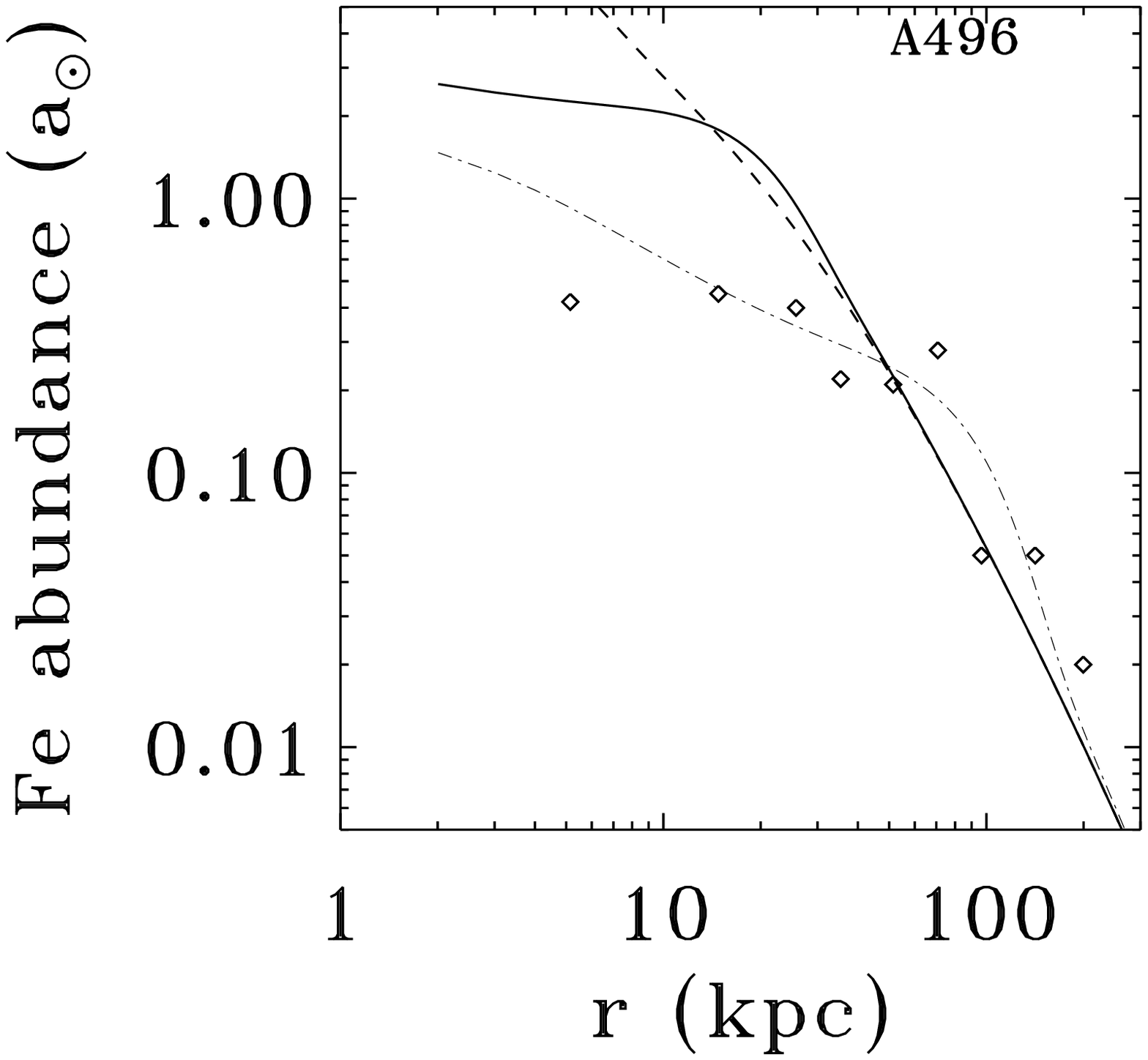}     \\      
    \includegraphics[width=0.45\hsize]{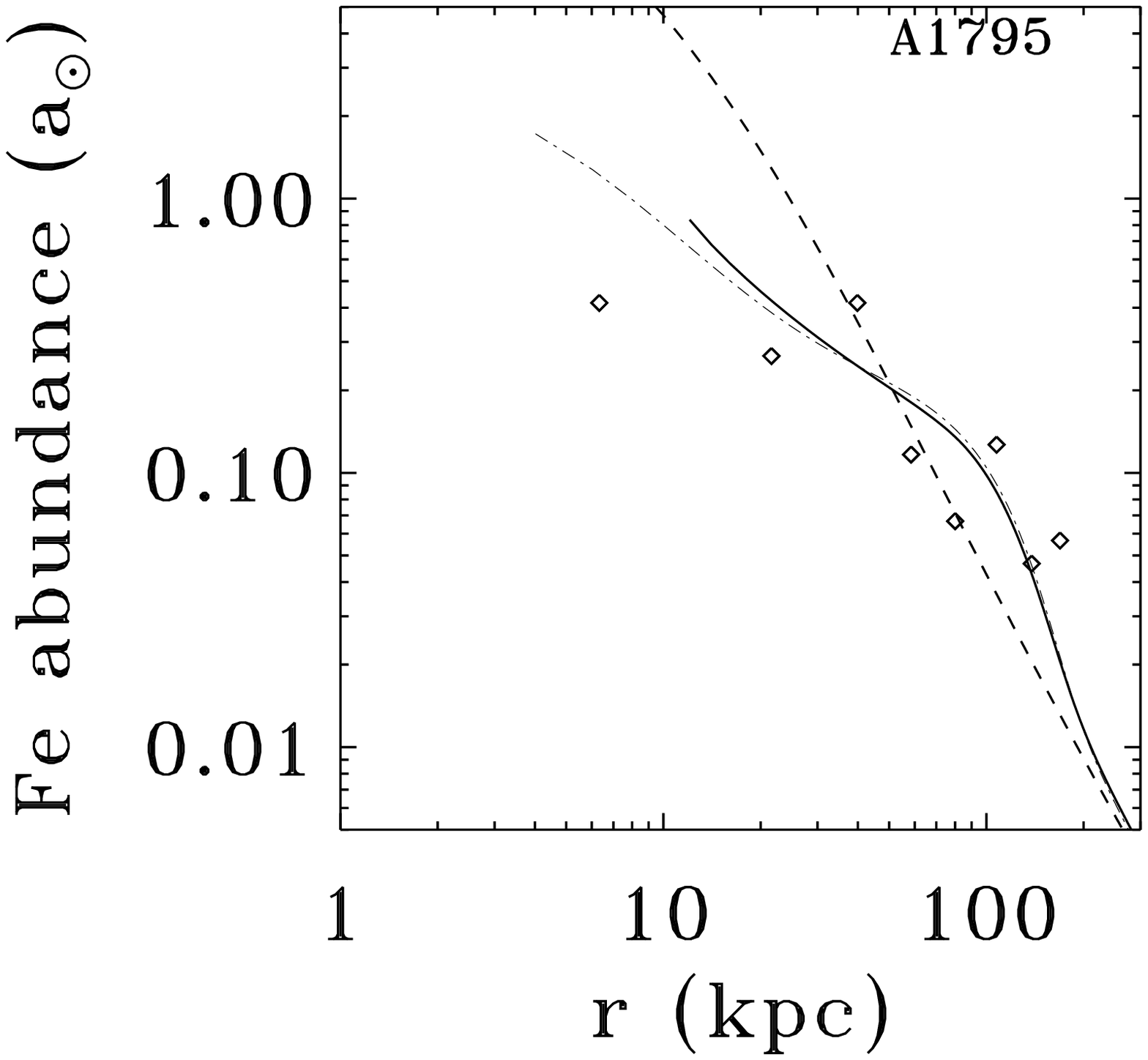}
    \includegraphics[width=0.45\hsize]{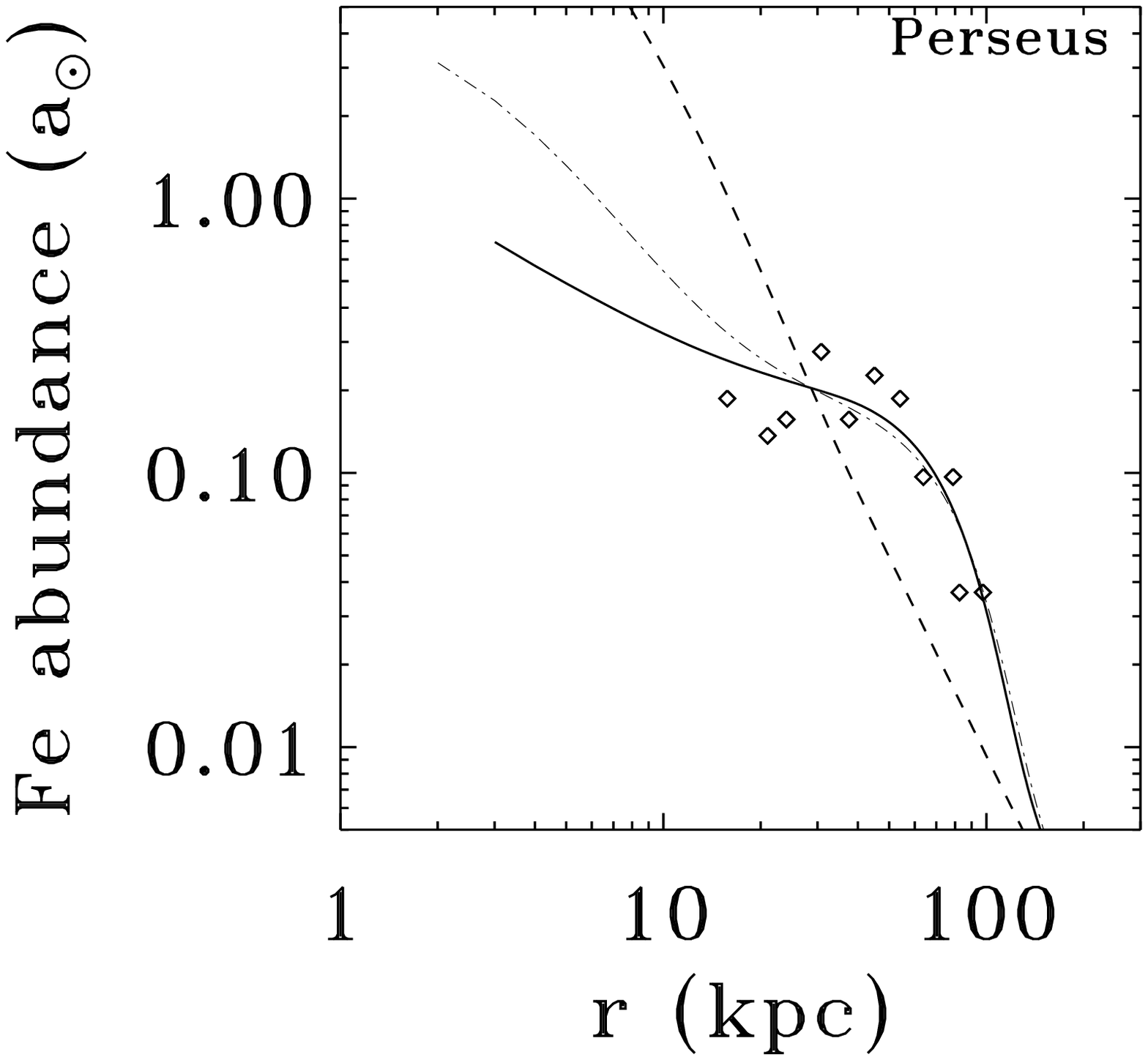}
  \end{tabular}  
  \caption{Abundance profile in solar unit as a function of the radius in kpc for Hydra A, Abell 496, Abell 1795, and Perseus. The dashed line is the resulting abundance profile without diffusion. The continuous line is the resulting profile with the present cluster diffusion. The dot-dashed line is the profile with the best fit cosmic-ray luminosity value and subsequent diffusion. The diamonds are the observations. Past and present AGN-driven convection seems to play an important role for the transport of iron. \label{result2}}
\end{figure*}

Using the model described in sections~\ref{sec:model_inj}
and~\ref{sec:model_diff} and the parameters from section \ref{sample},
we compute numerically the iron abundance profiles for the eight
clusters in our sample and compare to observations. For the sake of
comparison, we first compute the abundance profile without any
diffusion or inflow velocity. The result is shown as a dashed line in
Fig.\ref{result1} and Fig.\ref{result2}. Clearly, such profiles are
too steep near the center and do not match the observations
(diamonds). These profiles basically show that the observed
distributions of iron do not follow the distributions of stars within
the BCG galaxies. This is because the effective radii of the luminosity distribution $r_e$
(table \ref{data}) are smaller than the observed half iron mass radii
$r_{1/2}$ (table \ref{data2}).

Taking into account the diffusion and the inflow velocity brings the
model profiles substantially closer to the observations. This is
illustrated by the continuous lines in Fig.\ref{result1} and
Fig.\ref{result2}.  Interestingly, we find the same classification as
in subsection \ref{diffusion} when studying the intensity of the
diffusion coefficient. In the very ``weak feedback clusters'' with
the least diffusion (Virgo and Abell 496), abundances are largely
overestimated in the inner 20~kpc by a factor larger than four and
underestimated at $r>50$~kpc. ``Very weak feedback clusters'' do not
have enough present convection to diffuse the iron to larger radii. In
the ``weak feedback clusters'' with little convection (Abell 4059 and
Abell 262), abundances are underestimated near 30 kpc by a reasonable
factor of 1.5-2. The ``moderate feedback clusters'' with intermediate
diffusion coefficients (Perseus and Abell 1795) provide a much better
agreement with the observed abundance profiles. Their present
convection is almost enough to shape the abundance profiles so that
they are compatible with the observational points. Finally, in the
``strong feedback clusters'' (Sersic 159-03 and Hydra A), the
diffusion coefficients are sufficiently large or even too large
and the abundance
near the center is compatible (Sersic 159-03) or even too small by a
factor of $\sim 2$ in the inner $70$~kpc (Hydra A). We draw the
conclusion that turbulence at the level predicted by the AGN-driven convection model cannot be neglected when considering
the iron distribution in ``moderate'' or ``strong feedback
clusters''.

If AGN-driven convection is the dominant mechanism for transporting
metals, our iron abundance profile plots in Fig.\ref{result1} and
Fig.\ref{result2} suggest that Virgo-like clusters with low present
convection require a lot more convection in the past to match
observations whereas Hydra A-like clusters with a lot of present
convection require less past convection to match observations. This
coincides with the natural idea that clusters and more specifically
their central AGN were not in their current state during all their
history \citep{pope07}, an idea that we investigate
more thoroughly in the next section.

\section{Insight into the past variations of the AGN power from the shape of the abundance profiles}
\label{sec:results2} 

\citet{pope06} suggested that clusters are heated by Hydra-A type
events interspersed between epochs of lower AGN power. In this section
we explore this idea and the more general possibility that the AGN
power varies in some fashion during the lifetime of a cluster. As a
consequence of this variation, the amplitude of the AGN-driven
convection will also evolve.  Ideally, to investigate the effects of a
time-varying diffusion coefficient on the present-day abundance
profile, we would solve equation~(\ref{eqdiff}) using the correct
individual history of convection for each cluster in our sample.
However, we have no reliable way of determining these individual
histories. Instead, we adopt the following approach.  First, we
develop a model for relating the turbulent velocity profile in a
cluster directly to its AGN power (Sect.\ref{fitpart}). Second, for
each cluster in our sample, we determine the constant AGN power that
would be needed in order for the resulting turbulent velocity profile
to bring the iron abundance profile into agreement with observations.
We then take this constant AGN power to be a proxy for the
time-averaged AGN power within the cluster over the last $\sim
10$~Gyr.  In following this procedure, we make the strong hypothesis
that the AGN-driven convection is the only physical process that
transports iron in the ICM.

One way of thinking about the cause of the varying accretion rate of
the central AGN is provided by the same AGN-driven convection model of
\citet{chandran07} that we have been using to model the
turbulent velocity profile. In this model, the value of the AGN power
(which we take to be comparable to $L_{\rm cr}$) is determined
primarily by two factors: the density $\rho_{\rm core}$ and
temperature~$T_{\rm core}$ of the intracluster plasma at the radius
$r_{\rm core} \simeq 50-100$~kpc.  These parameters provide the outer
boundary conditions for the strongly cooling cluster core. For
larger~$\rho_{\rm core}$, the rate of radiative cooling within~$r_{\rm
  core}$ is larger, and thus more heating is needed to balance
cooling, leading to a larger value of~$L_{\rm cr}$. (In a
time-dependent model, if cooling were to initially exceed the total
heating, then the plasma would cool and flow inward, causing the AGN
mass accretion rate to grow until a balance between heating and
cooling is reached.)  Similarly, for a smaller~$T_{\rm core}$, the
conductive heating becomes smaller (the conductivity being $\propto
T^{5/2}$) and more convective heating is needed in order for total
heating to balance cooling.  In the AGN-driven convection model, small
variations in $\rho_{\rm core}$ and $T_{\rm core}$ cause large
variations in the AGN power.  The value of $L_{\rm cr}$ will thus
evolve as $\rho_{\rm core}$ and $T_{\rm core}$ change in time due to,
e.g., mergers, streams of infalling baryons, and stripping of gas from
cluster galaxies.

\subsection{Analytical fit of the convection profile depending on the AGN
  power only \label{fitpart}}


\begin{figure}[h!]
  \plotone{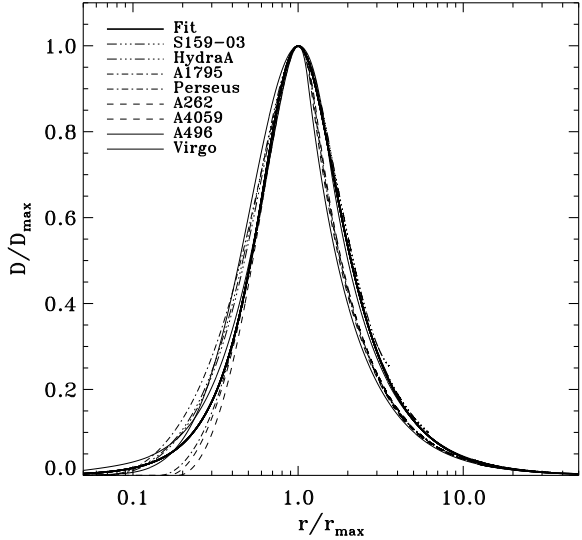}
  \caption{Diffusion coefficient (deduced from turbulent velocity field) normalized to its maximal value as a function of the radius normalized to the radius where the diffusion coefficient is maximum. The dashed lines are for our 8 clusters. Their profiles are self-similar. The thick continuous line is the fit given by Eq.\ref{eqfit}. \label{fit}}
\end{figure}

In Fig.\ref{fit} we plot $D/D_{\rm max}$ as a function of $r/r_{\rm max}$
for the eight clusters in our sample, where $r_{\rm max}$ is the
radius at which $D$ achieves its maximum. The shapes of the different
curves are similar, and so we are able to obtain a reasonable fit
to the diffusion coefficients using the anlalytic function
\begin{eqnarray}
\frac{D(r)}{D_{max}}&=&1.88 \times \left[1+\left(\frac{0.934r_{max}}{r}\right)^4\right] ^{-0.525}{}\nonumber\\ 
 {}&\times& \left[1+\left(\frac{r}{0.934 r_{max}}\right)^4\right]^{-0.4}.\label{eqfit}
\end{eqnarray}
The result of this fitting procedure is shown as a solid thick line in
Fig.\ref{fit}. For $r \gg r_{rmax}$ the right-hand side of
equation~(\ref{eqfit}) scales like $r^{-1.6}$ whereas for $r \ll
r_{max}$ it scales like~$r^{2.1}$. Equation~(\ref{eqfit}) matches  the
diffusion coefficient at $r>0.2 r_{max}$ fairly well. For smaller radii,
Eq.\ref{eqfit} is still compatible but the diffusion coefficient from
the model drops more strongly for some clusters.

\begin{figure}[h!]
				    \plotone{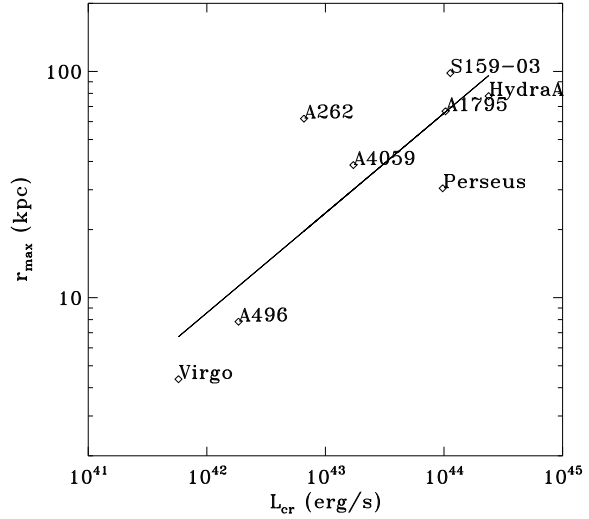}
  \caption{Correlation between the radius where the diffusion is maximum and the cosmic-ray luminosity. Right: Correlation between the maximum of the diffusion coefficient and the cosmic-ray luminosity.\label{correlationa}}
\end{figure}

\begin{figure}[h!]
				    \plotone{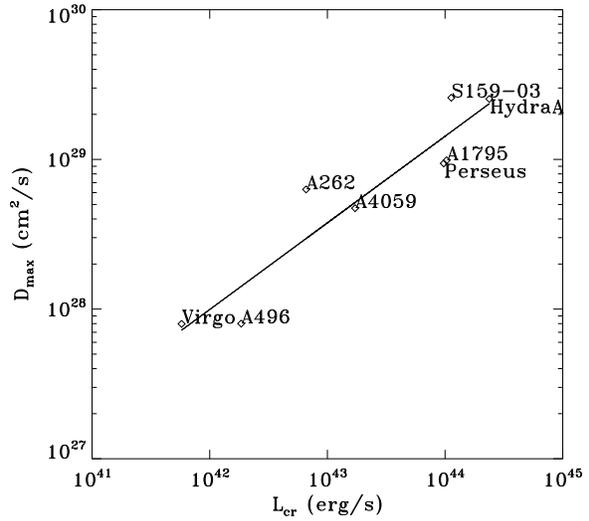}
  \caption{Correlation between the maximum of the diffusion coefficient and the cosmic-ray luminosity.\label{correlationb}}
\end{figure}

To develop a simpler one-parameter fit for $D(r)$, we note that
the quantities $D_{\rm max}$ and $r_{\rm max}$ are closely correlated
to the cosmic-ray luminosity~$L_{\rm cr}$
of the central AGN in the AGN-driven convection model.
These correlations are shown in Fig.\ref{correlationa}, Fig.\ref{correlationb},
and allow us to express $D_{\rm max}$ and $r_{\rm max}$
in terms of $L_{\rm cr}$ using the best-fit power-law relations:
\begin{eqnarray}
r_{max}&=&3.1 \times \left(\frac{L_{cr}}{10^{41}~\textrm{erg.s}^{-1}}\right)^{0.44}~\textrm{kpc}.
\label{eq:corr1}\\
D_{max}&=&2.6.10^{27}\left(\frac{L_{cr}}{10^{41}~\textrm{erg.s}^{-1}}\right)^{0.58}\!\!\!\! \textrm{cm}^2.\textrm{s}^{-1}
\label{eq:corr2} 
\end{eqnarray}
The correlation between $D_{\rm max}$ and $L_{\rm cr}$ results from the fact
that the convection is triggered by the gradient of
cosmic-ray pressure; thus, more cosmic-ray energy injection will
increase the convection level and the resulting diffusion. We note
that $L_{\rm cr}$ is not a free parameter of the AGN-driven convection
model, but rather is determined self-consistently within the model so
that the resulting convective heating (which increases with
increasing~$L_{\rm cr}$) is sufficiently large that the total heating
(convective plus conductive) balances radiative cooling.

\begin{figure}[h!]
  \plotone{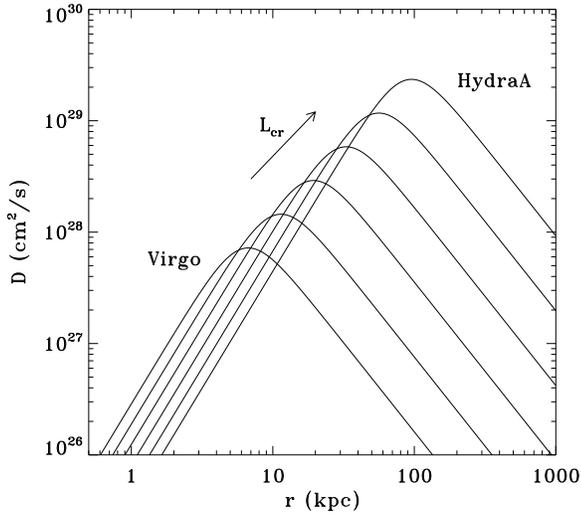}
  \caption{Diffusion coefficient profile (due to AGN-driven convection) as a function of the radius for different cosmic-ray  luminosities (AGN power). The cosmic-ray luminosities increases from $5.8\times 10^{41}~\textrm{erg.s}^{-1}$ (Virgo on the left) to $2.4 \times 10^{44}~\textrm{erg.s}^{-1}$ (Hydra A on the right) by a factor of $3.33$ each time. This comes from the analytical fit Eq.\ref{eqfit} and is used to determined which time average AGN power is required to match the observed abundance profiles. \label{plotdfit}}
\end{figure}

The accuracy of the analytical fit is of course degraded when
we use equations~(\ref{eq:corr1}) and (\ref{eq:corr2})  in
equation~(\ref{eqfit}), since we have
decreased the number of parameters. However, we have now achieved a
reasonable fit of the diffusion profile (due to AGN-driven convection)
depending only on~$L_{cr}$ (the
cosmic-ray luminosity or AGN power). The evolution of the profile as a
function of $L_{cr}$ is shown Fig.\ref{plotdfit} where the cosmic ray
luminosity is increased from Virgo's value ($L_{cr}=5.8\times 10^{41}
\textrm{erg.s}^{-1}$) to the Hydra A's value ($L_{cr}=2.4\times 10^{44}
\textrm{erg.s}^{-1}$) by a factor $3.33$ each time.

The diffusion term in Eq.\ref{eqdiff} can therefore be expressed as
a function of the cosmic-ray luminosity. We note that the inflow term in  Eq.\ref{eqdiff}
 can also be evaluated from the cosmic-ray luminosity since the accretion rate
 is given by
\begin{eqnarray}
\dot{M}&=&\frac{L_{cr}}{0.005 \times c^2}.
\label{eq:bondi} 
\end{eqnarray}
We have taken here the efficiency from \citet{chandran07}. This accretion rate
will be useful to solve Eq.\ref{eqdiff} in the next part.


\subsection{AGN power required to match the abundance profile \label{powerpart}}

\begin{figure}[h!]
  \plotone{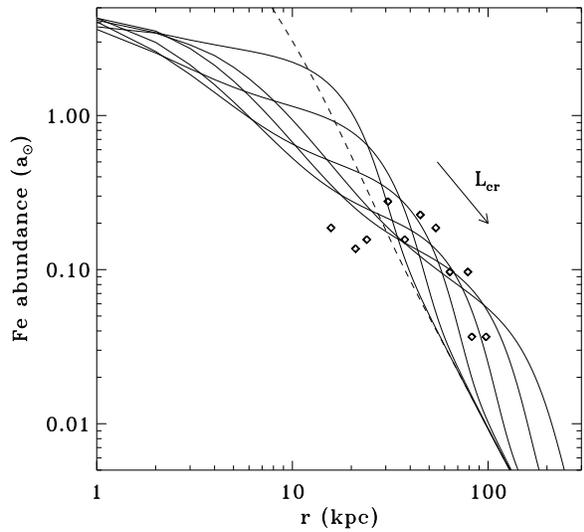}
  \caption{Perseus abundance profile as a function of the radius for different cosmic-ray  luminosities (AGN power). The cosmic-ray luminosities increases from $5.8\times 10^{41}~\textrm{erg.s}^{-1}$ (Virgo on the middle top) to $2.4 \times 10^{44}~\textrm{erg.s}^{-1}$ (Hydra A on the bottom right) by a factor of $3.33$ each time. It shows that by varying $L_{cr}$ we can reasonably match the abundance profile.\label{abundanceoflcr}}
\end{figure}

We now solve our model using the convection profile as given by
Eq.\ref{eqfit} (for the diffusion term) and the accretion rate as given by
Eq.\ref{eq:bondi} (for the inflow term). The only free parameter is the AGN power. We show in
Fig.\ref{abundanceoflcr} the evolution of the abundance profile as a
function of the AGN power. We deduce the luminosity $L_{cr}^{fit}$
which provides the best fit to the abundance profile. The resulting
profiles are presented in Fig.\ref{result1} and Fig.\ref{result2} as a
dashed-dotted line. The first conclusion is that we obtain a good
match for all the profiles by varying this single free parameter. This
was not obvious because we cannot adjust the size and the
normalization of the diffusion coefficient independently. Moreover,
the diffusion drops by a factor 10 outside of the region $0.2
r_{max}<r< 5 r_{max}$ and has very little effect outside this
region. The position of the peak of diffusion exactly where we need
diffusion to explain the abundance profile is a striking
coincidence. It means that by choosing the right AGN power one could
reproduce the shape of the observed iron excesses. We also compute the
half iron mass radius $r^{fit}_{1/2}$ for each
cluster (table \ref{dataout}). This is the radius which contains half the
total iron mass of the model inside $r_b$ (distance to the center of the last
observational point). The reasonable agreement with the observed half iron mass radius
$r_{1/2}$ (table \ref{data2}) computed inside the same radius $r_b$
confirms the above conclusion.

The following question naturally arises: are these average cosmic-ray
luminosities ($L_{cr}^{fit}$) realistic? Nothing guarantees realistic
values in our fitting process. The question is particularly
interesting for the ``very-weak feedback clusters'' such as Virgo and
Abell 262, and the ``strong feedback cluster'' Hydra A, which do not
fit the observed abundance profiles with the present convection.

The best-fit values of the cosmic-ray luminosity $L_{cr}^{fit}$ are
summarized in table \ref{dataout} and can be compared with the present
cosmic-ray luminosity $L_{cr}$ in table \ref{data}. Also shown in
table \ref{dataout} is the maximum of the diffusion coefficient
$D_{max}^{fit}$, and its radius $r_{max}^{fit}$, which can be compared
with the present value $r_{max}$ and $D_{max}$ in table \ref{data2}.
We recover here the same kind of value for the diffusion coefficient
value $D_{max}^{fit}=10^{28}-10^{29}~\textrm{cm}^2.\textrm{s}^{-1}$
(equivalent to maximum rms turbulent velocities between 70 and
110~km.s$^{-1}$) as \citet{rebusco05,rebusco06, graham06}. The
advantage in our approach is that the shape of the diffusion
coefficient comes from the physics-based AGN-driven model of
\citet{chandran07} and that we are also able to derive the
corresponding cosmic-ray luminosities.

\begin{figure}[h!]
  \plotone{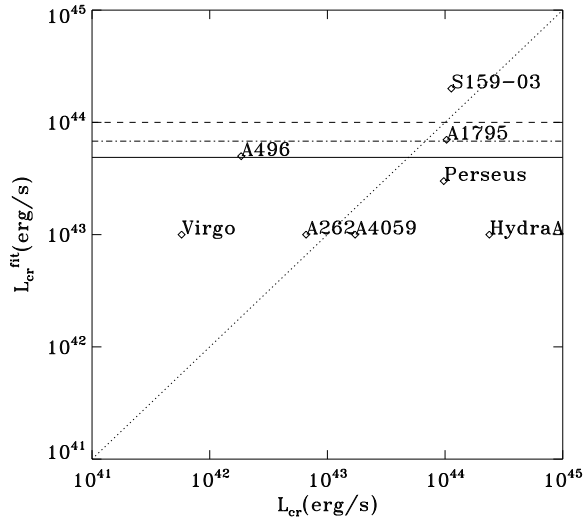}
  \caption{AGN power (cosmic-ray luminosity $L_{cr}^{fit}$) required to match
  the abundance profile as a function of the present AGN power (cosmic-ray
  luminosity $L_{cr}$) for our 8 clusters. The dotted line shows the equality
  between these two quantities. On this line, convection at present level
  sustained during the past 10~Gyr would be enough to explain the distribution of metals. The continuous line is the average $<L_{cr}^{fit}>= 4.9 \times 10^{43}$~erg.s$^{-1}$ over all the clusters. The dashed line is the estimate from \citet{voit05} using entropy-floor observations and the dot-dashed line is an estimate from \citet{best07} using radio observations. The proximity of these values to our average suggests an alternation of Virgo-type feedback, Perseus-type feedback, and Hydra-A type feedback and could explain the distribution of metals in clusters of galaxies.  \label{lcrfitplot}}
\end{figure}

Fig.\ref{lcrfitplot} shows the average cosmic-ray luminosities $L_{cr}^{fit}$
as a function of the present cosmic-ray luminosity $L_{cr}$. The dotted line
shows the equality between these two quantities. Clusters located near this
diagonal line have about the right amount of present convection to explain the
shape of the iron distribution (if this convection was sustained during the
past 10 Gyr). We recover the previously mentioned result that Hydra A has too
much convection whereas Virgo and A496 have too little (and the other clusters
have approximatively the right amount of present convection). What is
interesting is that the average $<L^{fit}_{cr}>=5 \times
10^{43}~\textrm{erg.s}^{-1}$ is about the same as the average of the present
cosmic-ray luminosities $<L_{cr}>=7 \times 10^{43}~\textrm{erg.s}^{-1}$. Moreover, the dispersion of cosmic-ray luminosities is much less since $L_{cr}^{fit}=10^{43}-10^{44}~\textrm{erg.s}^{-1}$ whereas $L_{cr}=10^{41}-10^{44}~\textrm{erg.s}^{-1}$. This is exactly what we would expect if in the cluster history the cosmic-ray luminosities evolved randomly between Virgo values, Perseus values and Hydra-A values. The average over all the clusters would be conserved and the dispersion of cosmic-ray luminosities would decrease.

We have searched the literature to find other constraints on the
average cosmic-ray luminosities in order to check if such values are
realistic. Direct measurement of the convection profile is, of course,
out of reach. The better constraints come from feedback
considerations. A study by \citet{pope06} of the heating rate of
cooling-core clusters suggests an alternation between Hydra-A type
events and smaller-scale outflows. By studying the entropy floor in
cooling-core clusters with and without radio emission, \citet{voit05,
  donahue06} suggest also an episodic heating. The invoked values are
$10^{45}$~erg.s$^{-1}$ for $10^7$~yr once every $10^{8}$~yr. This
leads to an average of $10^{44}$~erg.s$^{-1}$, which is comparable to
our value given the uncertainties.\\

The average of the observational estimates of the present mechanical
luminosities in a large sample of clusters could also provide a rough
estimate of the temporal average of the AGN power (although the
accuracy of such an estimate is reduced by its failure to take into
account evolution with redshift). \citet{birzan04,rafferty06} obtain a
sample of 33 cooling-core clusters with measured mechanical
luminosities. The values range from $10^{42}$~erg.s$^{-1}$ to
$10^{46}$~erg.s$^{-1}$ with an average of $9\times
10^{44}$~erg.s$^{-1}$ but we note that the sample is certainly biased
towards the high rate for $z>0.1$. If we remove these clusters, the
average falls to $2\times 10^{44}$~erg.s$^{-1}$. These samples contain
only a small number of clusters because of the difficulty of detecting
X-ray cavities and measuring their properties.\\ 

Another approach was taken by \citet{best07} who measured the radio
luminosities for the BCG of 625 nearby groups and clusters. Using the
conversion between mechanical luminosity and radio-luminosity from
\citet{birzan04,rafferty06}, they converted the radio luminosity
function into a mechanical luminosity function. The average over the
sample of BCG provides an estimate of the time-averaged mechanical
luminosities, $<L_{mech}>=2.3 \times 10^{42} f
(M_*/10^{11}~\textrm{M}_{\odot})~\textrm{erg.s}^{-1}$, with $f$ their
uncertainty factor, and $M_{*}$ the stellar mass. Assuming a
stellar mass to blue light ratio of 5.3 \citep{borriello03}, the
average of our stellar masses is $7.4\times
10^{11}~\textrm{M}_{\odot}$. A factor $f=1$ would mean that an energy
of $1~PV$ is considered per bubble. If we take $f=\gamma/(\gamma -1)=4$,
as in \citet{best07}, then we obtain an average mechanical power of
$<L_{mech}>=6.8 \times 10^{43}~\textrm{erg.s}^{-1}$, which is in good
agreement with our estimate using the iron distribution.\\ 

Another constraint comes from the mass of the central black hole,
which equals the time integral of the mass accretion rate
(plus a tiny contribution from the initial black hole seed plus a
possible contribution from black-hole mergers). Since the central
black hole may have grown substantially before the formation of the
cluster, the present value of the black hole mass provides only
an upper bound on the average mass accretion rate of the black
hole during the lifetime of the cluster.
Assuming a black hole of mass $3\times 10^{9}~\textrm{M}_{\odot}$ (as 
in Virgo), the average accretion rate must be smaller than
$0.3~\textrm{M}_{\odot}.\textrm{yr}^{-1}$.
This accretion rate corresponds to an average
cosmic-ray luminosity of $1.9 \times 10^{45} (\eta/0.1)
$~erg.s$^{-1}$, where $\eta$ is the efficiency with which the central
AGN converts the rest mass energy of accreted plasma into
cosmic-ray luminosity (which we take to be comparable to the total
AGN power). \citet{allen06} suggest that roughly $2.2~\%$ of the
Bondi accretion power is transformed into AGN mechanical luminosity.
Using this value for~$\eta$,
we find that the cosmic-ray luminosity must be less than roughly $4.1
\times 10^{44}$~erg.s$^{-1}$, which is larger than our
value. Moreover, it should be noted that the Bondi accretion rate may
significantly exceed the rate at which mass accretes onto the central
black hole, since much of the accreting plasma may end up forming
stars \citep{tan05} or being ejected in an outflow \citep{blan99}.
We thus conclude that our average value of $L_{cr}^{fit}$ is compatible
with present-day black hole masses.
Cosmological simulations from \citet{sijacki07,dimatteo07}
also indicate black hole accretion rates and mechanical luminosities
in the same range of values with large variations. In these
simulations, the final black hole mass comes from mergers and
accretion in high redshift quasar phases.\\

Our value of $L_{cr}^{fit} \sim 5 \times 10^{43}~\textrm{erg.s}^{-1}$
seems reasonable and compatible with current constraints (see
Fig.\ref{lcrfitplot}). If our scenario is correct, it indicates that
the iron distribution could put strong constraints on the AGN feedback
model. For instance, an average AGN power greater than $2.4 \times
10^{44}~\textrm{erg.s}^{-1}$ would destroy the observed abundance
profile of most of the clusters as illustrated
Fig.\ref{abundanceoflcr}. Moreover, our scenario gives further support
for the idea that convection plays a fundamental role in clusters of
galaxies by explaining the density and temperature profile as well as
the shape of the iron abundance excess. Our model is, however, very
simplistic and relies on many approximations. In order to be more
realistic, one would have to carry out cosmological simulations with
AGN feedback and supernov{\ae} enrichment as
\citet{sijacki07,dimatteo07}. However, a proper treatment of the
convection would also require the inclusion of cosmic rays and anisotropic
transport \citep{rasera08} since these two ingredients modify the
convective instability criterion \citep{balbus00,chandran06,dennis08,parrish08,quataert08}.

\section{Conclusion}
\label{sec:conclusion}

In this article, we have studied the impact of AGN-driven convection
on the shape of the iron abundance profile for 8 cooling-core
clusters: Perseus, Hydra A, Sersic 159-03, Abell 262, Abell 1795,
Virgo, Abell 496 and Abell 4059. We have used the iron injection model
from \citet{bohringer04} where metal production by SNIa and winds
follows the Hernquist light profile of the central brightest cluster galaxy (BCG). We have
also used the steady-state convection model from \citet{chandran07} to
determine the turbulent velocity profile in the ICM.  In this model,
AGN-driven convection is the dominant mechanism for transferring
energy from the central AGN to the ICM thereby preventing large
quantities of plasma from cooling to low temperatures.\\

Stochastic motions with an rms radial velocity fluctuation $u_{NL}$ correspond
to a diffusion coefficient $D=0.5 l u_{NL}$ in the mixing length-theory
employed by \citet{chandran07}, where
$l$ is the mixing-length which we set to $0.4 r$. We have solved a 1D
advection-diffusion equation for the iron abundance
profile, and compared our results to XMM and Chandra abundance profiles. The
profiles obtained without diffusion (metal injection only) are too peaked
toward the center. Taking into account AGN-driven convection improves the
abundance profile because it smoothes the center. For most objects in our
sample the modeled profiles do not differ greatly from the observations.
However, the less convective clusters (Abell 496 and Virgo) require
much more convection in order to match observations
whereas the most convective cluster (Hydra A)
requires less.\\

Making use of the approximate self-similarity of the diffusion
coefficient profiles in the AGN-driven convection model, we model the
diffusion coefficient as a function of radius~$r$ that depends only on
one parameter, namely the cosmic-ray luminosity (AGN power). We have
used this function to find if any reasonable AGN power could provide a
good fit to the observed iron distributions.  Although the diffusion
coefficient is non-negligible only in a limited range of radii around
its maximum, and although we cannot adjust the amplitude and the
position of this maximum independently, we find a good match for all
the clusters.\\

The best-fit AGN powers $L_{cr}^{fit}$ can be thought as the time
average over the cluster history (10~Gyr) required to explain entirely
the width of the abundance profile. We found a range of values for
$L_{cr}^{fit}$ between $10^{43}$ and $2\times
10^{44}~\textrm{erg.s}^{-1}$, with an average of $5 \times
10^{43}~\textrm{erg.s}^{-1}$. Such a value seems quite reasonable and
compatible with current constraints on the entropy profile, black hole
mass, and radio-inferred average mechanical luminosity. The shape of
the abundance profile could therefore be a fossil record of the past
AGN-driven convection.

This model relies on several approximations. The separation of the
abundance profile into a constant component $a_{b}$ (from the
background galaxies and SNII) and an iron excess (from the central BCG) 
is one of the major approximations. A better method would be
to also include the contribution from other galaxies and SNII and to
compare with the total abundance profile, however this contribution
also suffers from a lot of uncertainties. We also neglect all other
possible sources of convection such as supernova winds, the jet
itself, and stirring by infalling galaxies. This may explain the
discrepancy between our results and the observed abundances at the
very centers of several of the clusters, in some of which
an abundance hole is observed. Another step towards a more
realistic model would be cluster simulations with metal injection and
AGN feedback \citep{sijacki07,roediger07} but also cosmic-ray
injection, anisotropic transport and cosmic-ray diffusion. By
modifying the convective instability criterion these three last
ingredients should drive turbulence to a level close to the one
predicted by the mixing length theory.

\acknowledgments This work was partially supported by NASA's Astrophysical Theory Program under grant NNG 05GH39G and by NSF under grant AST 05-49577. We acknowledge the usage of the HyperLeda database (http://leda.univ-lyon1.fr).

\clearpage


\clearpage

\begin{table*}[h!]
\begin{center}
\begin{tabular}{|c|c|c|c|c|c|c|}

\hline
Cluster&z&M$_{vir}$&L$_{cr}$&BCG&L$_B$&r$_e$\\
&&(h$^{-1}$M$_{\odot}$)&(erg.s$^{-1}$)&&(L$^B_{\odot}$) &(kpc)\\
\hline
Virgo&0.00372&$3.1\times 10^{14}$&$5.8 \times 10^{41}$&M87&$6.4 \times 10^{10}$&6.1\\
\hline
Abell262&0.0155&$3.2\times 10^{13}$&$6.6 \times 10^{42}$&NGC708&$3.8 \times 10^{10}$&24\\
\hline
Sersic159-03&0.0572&$9.6\times 10^{13}$&$1.1 \times 10^{44}$&ESO291-009&$1.3 \times 10^{11}$&27\\
\hline
Abell4059&0.0466&$6.8\times 10^{14}$&$1.7 \times 10^{43}$&ESO349-010&$1.9 \times 10^{11}$&22\\
\hline
HydraA&0.0550&$7.1\times 10^{13}$&$2.4 \times 10^{44}$&3C218&$2.6 \times 10^{11}$&44\\
\hline
Abell496&0.0322&$8.2 \times 10^{13}$&$1.9 \times 10^{42}$&PGC015524&$1.6 \times 10^{11}$&50\\
\hline
Abell1795&0.0639&$5.0\times 10^{14}$&$1.0 \times 10^{44}$&PGC049005&$1.1 \times 10^{11}$&42\\
\hline
Perseus&0.0179&$6.2\times 10^{14}$&$9.8 \times 10^{43}$&NGC1275&$1.7 \times 10^{11}$&15\\
\hline
\end{tabular}
\caption{Clusters and BCGs general properties \label{data}}
 \tablecomments{z is the redshift, $M_{vir}$ is the virial mass, $L_{cr}$ is the cosmic-ray luminosity (AGN power), BCG is the brightest cluster galaxy, $L_{B}$ is the blue luminosity of the BCG and $r_{e}$ is the effective radius which contains half the projected luminosity. Methods, definitions and references are described in section \ref{sample}.}
\end{center}
\end{table*}

\begin{table*}[h!]
\begin{center}
\begin{tabular}{|c|c|c|c|c|c|c|}

\hline
Cluster&$r_{max}$&$D_{max}$&$r_{b}$&$a_b$&$M_{Fe}$&$r_{1/2}$\\
&(kpc)&(cm$^2$.s$^{-1}$)&(kpc)&(a$_\odot$)&(M$_{\odot}$)&(kpc)\\
\hline
Virgo&4.4&$8.0 \times 10^{27}$&58&0.20&$4.2 \times 10^7$&36\\
\hline
Abell262&62&$6.3 \times 10^{28}$&200&0.29&$1.4 \times 10^8$&55\\
\hline
Sersic159-03&98&$2.6 \times 10^{29}$&260&0.14&$2.4 \times 10^8$&110\\
\hline
Abell4059&39&$4.7 \times 10^{28}$&85&0.43&$1.9 \times 10^8$&44\\
\hline
HydraA&78&$2.5 \times 10^{29}$&140&0.28&$2.0 \times 10^8$&57\\
\hline
Abell496&7.8&$8.0 \times 10^{27}$&200&0.28&$2.6 \times 10^8$&80\\
\hline
Abell1795&67&$9.9 \times 10^{28}$&170&0.33&$4.8 \times 10^8$&92\\
\hline
Perseus&30&$9.4 \times 10^{28}$&98&0.46&$1.7 \times 10^8$&53\\
\hline
\end{tabular}
\caption{Diffusion properties derived in the AGN-convection model and abundance profile parameters \label{data2}}
 \tablecomments{ $r_{max}$ is the position of the maximum of the diffusion
 coefficient and $D_{max}$ is the maximum. $r_{b}$ is the radius of the last
 bin of the observed abundance profile, $a_b$ is the background abundance
 (relative to the solar abundance of \citet{anders89}) which is subtracted to
 the abundance profile to obtain the iron excess, $M_{Fe}$ is the total
 observed iron mass, and $r_{1/2}$ is the iron half mass radius (which
 therefore contains an iron mass of $M_{Fe}/2$). Methods, definitions and references are described in section \ref{sample}.}
\end{center}
\end{table*}

\begin{table*}[h!]
\begin{center}
\begin{tabular}{|c|c|c|c|c|c|}

\hline
Cluster&L$_{cr}^{fit}$&$r^{fit}_{max}$&$D^{fit}_{max}$&sr$_{eff}$&$r^{fit}_{1/2}$\\
&(erg.s$^{-1}$)&(kpc)&(cm$^2$.s$^{-1}$)&SNU&(kpc)\\
\hline
Virgo&$1.0 \times 10^{43}$&23&$3.8 \times 10^{28}$&0.082&33\\
\hline
Abell262&$1.0 \times 10^{43}$&24&$3.8 \times 10^{28}$&0.21&47\\
\hline
Sersic159-03&$2.0 \times 10^{44}$&87&$2.1 \times 10^{29}$&0.19&99\\
\hline
Abell4059&$1.0 \times 10^{43}$&23&$3.8 \times 10^{28}$&0.11&44\\
\hline
HydraA&$1.0 \times 10^{43}$&23&$3.8 \times 10^{28}$&0.082&54\\
\hline
Abell496&$5.0 \times 10^{43}$&48&$9.6 \times 10^{28}$&0.17&79\\
\hline
Abell1795&$7.0 \times 10^{43}$&55&$1.2 \times 10^{29}$&0.43&81\\
\hline
Perseus&$3.0 \times 10^{43}$&37&$7.1 \times 10^{28}$&0.11&51\\
\hline
\end{tabular}
\caption{AGN power and diffusion properties for the best fit model to the
 abundance profile \label{dataout}}
 \tablecomments{ $L_{cr}^{fit}$ is the cosmic-ray luminosity (AGN power)
 required to reproduce the observed abundance profile. From this AGN power, we
 derived the diffusion coefficient profile. $r_{fit}^{max}$ is the
 radius of the maximum of the diffusion coefficient profile and
 $D_{fit}^{max}$ is the maximum. $sr_{eff}$ is the effective SNIa rate (which
 includes the wind contribution to the iron injection). $r^{fit}_{1/2}$ is the
 iron half mass radius of the model (which contains half the total iron mass
 inside $r_b$). Methods and definitions are described in section \ref{model} and section \ref{sec:results2}.}
\end{center}
\end{table*}

\end{document}